\documentclass[12pt, preprint]{aastex}
\usepackage{natbib}
\usepackage[american]{babel}
\usepackage{mathpazo}
\usepackage{color}

\usepackage{amssymb,amsmath,wasysym}

\usepackage{multirow}

\newcommand{\bit}{\begin{itemize}}
\newcommand{\eit}{\end{itemize}}
\newcommand{\bbl}{\begin{block}{}}
\newcommand{\ebl}{\end{block}}
\newcommand{\babl}{\begin{alertblock}{}}
\newcommand{\eabl}{\end{alertblock}}

\newcommand{\Eth}{ {\ensuremath{E_{\textrm{min}}}} }
\newcommand{\sigmav}{ {\ensuremath{\langle \sigma v \rangle}} }

\newcommand{\jbar}{ {\ensuremath{\overline{J}} }}
\newcommand{\jbardo}{ {\ensuremath{\overline{J}(\Delta\Omega)} }}

\newcommand{\gr}{{\ensuremath{\gamma \textrm{-ray}} }}
\newcommand{\grs}{{\ensuremath{\gamma \textrm{-rays}} }}
\newcommand{\grad}{\ensuremath{^{\textsf{o}}}}
\newcommand{\gev}{ {\ensuremath{\textrm{ GeV}} }}

\newcommand{\msun}{{\ensuremath{{\textrm M}_{\tiny\astrosun}}}}
\newcommand{\hess}{{H.E.S.S.} }
\newcommand{\dd}{\ensuremath{\mathrm{d}}}

\newcommand{\LL}{\ensuremath{\mathcal{L}} }

\begin{document}
\bibliographystyle{natbib}

\title{Search for Dark Matter Annihilation Signals from the Fornax Galaxy
  Cluster with H.E.S.S.}


\author{H.E.S.S. Collaboration:
A.~Abramowski\altaffilmark{1},
F.~Acero\altaffilmark{2},
F.~Aharonian\altaffilmark{3,4,5},
 A.G.~Akhperjanian\altaffilmark{6,5},
 G.~Anton\altaffilmark{7},
 A.~Balzer\altaffilmark{7},
 A.~Barnacka\altaffilmark{8,9},
 U.~Barres de Almeida\altaffilmark{10,\ast},
 Y.~Becherini\altaffilmark{11,12}
% ...
}

% ...
\altaffiltext{1}{Universit\"at Hamburg, Institut f\"ur Experimentalphysik, Luruper Chaussee 149, D 22761 Hamburg, Germany}
 \altaffiltext{2}{Laboratoire de Physique Th\'eorique et Astroparticules, Universit\'e Montpellier 2, CNRS/IN2P3, CC 70, Place Eug\`ene Bataillon, F-34095 Montpellier Cedex 5, France}
 \altaffiltext{3}{Max-Planck-Institut f\"ur Kernphysik, P.O. Box 103980, D 69029 Heidelberg, Germany}
\altaffiltext{4}{Dublin Institute for Advanced Studies, 31 Fitzwilliam Place, Dublin 2, Ireland}
\altaffiltext{5}{National Academy of Sciences of the Republic of Armenia, Yerevan}
\altaffiltext{6}{Yerevan Physics Institute, 2 Alikhanian Brothers St., 375036 Yerevan, Armenia}
\altaffiltext{7}{Universit\"at Erlangen-N\"urnberg, Physikalisches Institut, Erwin-Rommel-Str. 1, D 91058 Erlangen, Germany }
\altaffiltext{8}{Nicolaus Copernicus Astronomical Center, ul. Bartycka 18, 00-716 Warsaw, Poland}
\altaffiltext{9}{IRFU/DSM/CEA, CE Saclay, F-91191 Gif-sur-Yvette, Cedex, France}
\altaffiltext{10}{University of Durham, Department of Physics, South Road, Durham DH1 3LE, U.K.}
\altaffiltext{11}{Astroparticule et Cosmologie (APC), CNRS, Universit\'e Paris 7 Denis Diderot, 10,
rue Alice Domon et Leonie Duquet, F-75205 Paris Cedex 13, France.
 Also at UMR 7164 (CNRS, Universit\'e Paris VII, CEA, Observatoire de Paris)}
\altaffiltext{12}{Laboratoire Leprince-Ringuet, Ecole Polytechnique, CNRS/IN2P3,
 F-91128 Palaiseau, France}
% ...
\altaffiltext{$\ast$}{supported by CAPES Foundation, Ministry of Education of Brazil}
% ...


%
%




\begin{abstract}
The Fornax galaxy cluster was observed with the High Energy
Stereoscopic System (H.E.S.S.) for a total live time of 14.5 hours,
searching for very-high-energy (VHE, $E>100\gev$) \grs from dark
matter (DM) annihilation.
No significant signal was found in
searches for point-like and extended emissions. Using several models of the
DM density distribution, upper limits on the DM velocity-weighted
annihilation cross-section $\left\langle \sigma v \right\rangle$
as a function of the DM particle mass are
derived. Constraints are derived for different DM particle
models, such as those arising from Kaluza-Klein and supersymmetric
models. Various annihilation final states are considered. Possible
enhancements of the DM annihilation \gr flux, due to DM substructures of
the DM host halo, or from the \textit{Sommerfeld} effect, are studied. Additional
\gr contributions from internal bremsstrahlung and inverse Compton
radiation are also discussed. For a DM particle mass of
1 TeV, the exclusion limits at 95\% of confidence level reach values of $\langle \sigma v \rangle^{\rm  95\%\,C.L.} \sim$ 10$^{-23}$ cm$^3$s$^{-1}$, depending on the DM particle model and halo properties. Additional contribution
from DM substructures can improve the upper limits on $\langle \sigma v \rangle$ by more than two orders of magnitude. At masses around 4.5 TeV, the enhancement by substructures and the Sommerfeld resonance effect results in a velocity-weighted annihilation cross-section upper limit at the level of $\langle \sigma v \rangle^{\rm  95\%\,C.L.} \sim$10$^{-26}$ cm$^3$s$^{-1}$.
\end{abstract}




\section{Introduction}

Galaxy clusters are the largest virialized objects
observed in the Universe. Their main mass component is dark matter
(DM), making up about 80$\%$ of their total mass budget, with the
remainder provided by intracluster gas and galaxies, at 15$\%$ and
5$\%$ respectively~\citep[see e.g][]{2005RvMP...77..207V}. The DM halo distribution within galaxy
clusters appears to be well reproduced by N-body numerical simulations
for gravitational structure formation~\citep[and references therein]{2006AA...455...21C,2008A&A...478L..23R,2010AA...513A..52S,2005RvMP...77..207V}. This may be in contrast to smaller systems like dwarf galaxies. For instance,
disagreements between theoretical predictions and actual estimates of the DM halo profile from observations have been found in low surface brightness galaxies~\citep{1998ApJ...499...41M,1998astro.ph..7084N,2010AdAst2010E...5D}. Although such discrepancies may vanish at galaxy cluster scale, the influence of baryon infall in the DM gravitational potential can still flatten the DM density distribution in the inner regions of galaxy clusters~\citep[see, for instance,][]{2001ApJ...560..636E}.

The pair annihilation of weakly interacting massive particles (WIMP)
constituting the DM halo is predicted to be an important source of
non-thermal particles, including a significant fraction as photons
covering a broad multiwavelength spectrum of emission~\citep[see, for instance,][]{Bergstrom:2000pn,2006AA...455...21C}.
Despite the fact that galaxy clusters are located at much further distances than the  dwarf
spheroidal galaxies around the Milky Way, the higher annihilation luminosity of
clusters make them comparably good targets for indirect detection of dark matter.
The flux of \grs from WIMP DM annihilation in
clusters of galaxies is possibly large enough to be detected by current \gr
telescopes~\citep{2009PhRvD..80b3005J,2009PhRvL.103r1302P}.
Also standard astrophysical scenarios have been proposed for \gr
emission~\citep[see e.g.][for a review]{Blasi:2007pm}, in particular,
collisions of intergalactic cosmic rays and target nuclei from the intracluster
medium. Despite these predictions, no significant \gr emission has
been observed in local clusters by H.E.S.S.~\citep{Aharonian:2008uq,Aharonian:2009bc}, MAGIC~\citep{2010ApJ...710..634A} and \emph{Fermi}-LAT~\citep{2010JCAP...05..025A,2010ApJ...717L..71A} collaborations. However, \grs of a different astrophysical emission processes have already been detected from some central radio galaxies in clusters (e.g. \citet{2006Sci...314.1424A,Acciari:2008ah,2010ApJ...723L.207A,Abdo:2009ta}).


Following the absence of a signal, upper limits for a DM annihilation signal coming from galaxy clusters have been published by \emph{Fermi}-LAT~\citep{2010JCAP...05..025A} and MAGIC~\citep{2010ApJ...710..634A} collaborations.
Strong constraints on the annihilation cross-section of DM from the Fornax galaxy cluster have been put by the \emph{Fermi}-LAT collaboration for DM particles masses up to 1 TeV from \gr selected in the 100 MeV - 100 GeV energy range. However many DM models show distinct features in the DM annihilation spectrum close to DM particle mass, such as monochromatic gamma-ray lines, sharp steps or cut-offs, as well as pronounced bumps. This could provide a clear distinction between an annihilation signal and a standard astrophysical signal~\citep[see, for instance,][]{2011PhRvD..84j3525B}). These features are often referred as \textit{smoking-gun} signatures. Such models can only be tested by satellite telescopes for DM particle masses up to a few hundreds of GeV. IACTs observation can provide well-complementary searches for such features at DM particle masses higher than a few hundreds of \gev.

This paper reports on the observation in VHE $\gamma$ rays of the Fornax galaxy
cluster (ACO S373) with the High Energy
Stereoscopic System (H.E.S.S.). Interdependent constraints on several
DM properties are derived from the data, such as the DM particle mass
and annihilation cross-section. Different models of the DM density
distribution of the cluster halo are studied. The paper is structured
as follows. In Section 2 the Fornax galaxy cluster is described. The choice of Fornax for a DM analysis is
motivated, based on the DM content and distribution inside the
cluster.
Section 3 presents the data analysis and results. Upper limits on the \gr flux for both standard astrophysical sources and DM annihilation are extracted in Section 4. Exclusion limits on the DM annihilation
cross-section versus the particle mass are given in Section 5. Several DM particle
candidates are considered, with particular emphasis on possible particle
physics and astrophysical enhancements to the \gr annihilation flux.


\section{Target selection and dark matter content}
\label{sec:jbar}
The Fornax~\citep[distance = 19 Mpc,][]{Tonry:2000aa}, Coma~\citep[distance = 99 Mpc,][]{2002ApJ...567..716R} and Virgo~\citep[distance = 17 Mpc,][]{Mei:2007xs} galaxy clusters are in principle promising targets for
indirect dark matter searches through $\gr$s, as was shown by
\cite{2009PhRvD..80b3005J}. The radio galaxy M 87
at the center of Virgo provides a strong astrophysical \gr
signal ~\citep{2006Sci...314.1424A}, showing flux
variabilities from daily to yearly timescales that exclude the bulk of
the signal to be of a DM origin. Since a DM \gr signal would be hard
to disentangle from this dominant standard astrophysical
signal, Virgo is not a prime target for DM searches, even though a DM
signal may be hidden by the dominant \gr signal from standard astrophysical sources.

Moreover, galaxy clusters are
expected to harbor a significant population of relativistic cosmic-ray
protons originating from different sources, such as large-scale shocks
associated with accretion and merger
processes~\citep{1998APh.....9..227C,2003ApJ...593..599R},
 or supernovae~\citep{1996SSRv...75..279V} and AGN
activity~\citep{2007MNRAS.382..466H}. The \gr
emission arising from pion decays produced by the interaction of
these cosmic-ray protons with the intracluster gas may be a potential
astrophysical background to the DM-induced \gr signal. In the case of
Coma, \cite{2009PhRvD..80b3005J} showed that such astrophysical
background is expected to be higher than the DM annihilation
signal\footnote{Also the two brightest radio galaxies, NGC 4874 and NGC 4889, lying in the central region of Coma may be potential sources of a standard astrophysical \gr signal.}. On the other hand, the same study ranked Fornax as the most
luminous cluster in DM-induced \gr emission among a sample of 106 clusters
from the HIFLUGCS catalog~\citep{2002ApJ...567..716R}. The
DM-to-cosmic-ray \gr flux ratio of Fornax was predicted to be larger
than 100 in the GeV energy range~\citep{2009PhRvD..80b3005J}. A recent independent study by~\citet{Pinzke:2011ek} has also predicted Fornax to be among the brightest DM galaxy clusters with a favorably-low cosmic-ray induced signal.
Although the central galaxy of the Fornax cluster, NGC 1399, is a radio galaxy and could
in principle emit \grs, the super-massive black hole at the center of this galaxy have been shown to be passive~\citep{2011ApJ...738..142P}.
Indeed recent observations of several clusters with the \emph{Fermi}-LAT detector have shown no \gr
signal~\citep{2010ApJ...717L..71A}, and the most stringent limits on
dark matter annihilation were derived from the Fornax observations~\citep{2010JCAP...05..025A}.

The center of Fornax galaxy cluster is located at RA(J2000.0) =
$03^{\rm h}38^{\rm m}29^{\rm s}_{^{\cdot}}3$ and \\ Dec(J2000.0)~=~$-35^{\circ}$~$27^{\prime}$~$00^{\prime \prime}_{^{\cdot}}7$
 in the Southern Hemisphere. For ground-based Cherenkov telescopes
 like \hess (cf. Section \ref{sec:hess}),
low zenith angle observations are required to guarantee the lowest
possible energy threshold and the maximum sensitivity of the
instrument. Given the location of H.E.S.S., this
condition is best fulfilled for Fornax, compared to the Virgo and Coma clusters.
Therefore, Fornax is the preferred galaxy cluster target for dark
matter searches for the \hess experiment. The properties of its dark matter
halo are discussed in more details in the following section.

\subsection{Dark matter in the Fornax galaxy cluster}
\label{sec:fornaxdm}

The energy-differential \gr flux from dark matter annihilations is given by
the following equation:
\begin{equation}
  \frac{\dd \Phi_{\gamma}(\Delta\Omega,E_{\gamma})}{\dd E_{\gamma}}\,=\frac{1}{8\pi}\,
  \frac{\langle \sigma    v\rangle}{m_{\rm DM}^2}
  \,\frac{\dd N_{\gamma}}{\dd E_{\gamma}}
  \,\times\,\jbar(\Delta\Omega)\Delta\Omega \:,
\label{eq:flux}
\end{equation}
where \sigmav is the velocity-weighted annihilation cross-section,
$m_{\rm DM}$ the mass of the DM particle and $\dd N_{\gamma}/\dd E_{\gamma}$ the photon
spectrum per annihilation. The factor
\begin{equation}
  \overline{J}(\Delta \Omega) =
  \frac{1}{\Delta\Omega}\int_{\Delta\Omega} \dd \Omega
  \int_{\textrm{LOS}} \dd l \times \rho^2[r(l)]
\label{eq:jbar}
\end{equation}
reflects the dark matter density distribution
inside the observing angle $\Delta\Omega$. 
The annihilation luminosity scales with the squared dark matter
density  $\rho^2$, which is conveniently parametrized as a function of the radial
distance $r$ from the center of the astrophysical object under
consideration. This luminosity is integrated along the line of sight 
(LOS) and within an angular region $\Delta\Omega$, whose optimal
value depends on the dark matter profile of the target and the
angular resolution of the instrument.

Numerical simulations of structure formation in the $\Lambda$CDM
framework predict cuspy dark matter halos in galaxies and clusters of galaxies
\citep{1996ApJ...462..563N,1997ApJ...477L...9F,1998ApJ...499L...5M}. A
prominent parametrization of such halos is  the
``Navarro-Frenk-White'' (NFW) profile~\citep{Navarro:1996gj},
characterizing halos by their scale radius $r_s$ at which the
logarithmic slope is  $ \dd\ln \rho / \dd \ln r  = -2$, and a
characteristic density $\rho_s = 4\,\rho(r_s)$. This profile was shown to be consistent with X-ray observations of the intracluster medium of galaxy clusters. The DM density profile is
given by:

\begin{equation}
\rho_{\rm NFW}(r) =
\frac{\rho_s}{\left(\frac{r}{r_s}\right)\left(1+\frac{r}{r_s} \right)^2} \quad.
\label{eq:nfw}
\end{equation}
Another prediction of $\Lambda$CDM N-body simulations is an abundance of halo
substructures, as will be detailed in section \ref{sec:subs}. On the other hand, in scenarios where the baryon infall in the DM gravitational potential efficiently transfers energy to the inner part of the DM halo by dynamical friction, a flattening of the density cusp into a core-halo structure is predicted~\citep[see e.g.][]{2001ApJ...560..636E}. These halos can be parametrized by the ``Burkert profile'' ~\citep{Burkert:1995yz}:
\begin{equation}
\rho_B(r) = \frac{\rho_0 r_c^3}{(r+r_c)(r^2 + r_c^2)}\quad.
\label{eq:burk}
\end{equation}
Again, the dark matter density falls off as $\sim
r^{-3}$ outside the core radius $r_c$, but it approaches a constant
value $\rho_0$ for $r \rightarrow 0$. In the following, dark matter
halos of both types are considered.

A commonly-used approach for the determination of the DM halo in galaxy cluster comes from X-ray measurements of the gravitationally bound hot intracluster
gas. From the HIFLUGCS catalog~\citep{2002ApJ...567..716R}, the virial
mass and radius of Fornax  are found to be
$M_{\rm vir}\sim 10^{14}$ M$_{\odot}$ and $R_{\rm vir} \sim 1$ Mpc
(corresponding to about 6$^\circ$ in angular diameter), respectively.
Under the assumption of a NFW halo profile in $\Lambda CDM$ cosmology, a relation between the
virial mass and the concentration parameter $c=R_{\rm vir}/r_s$ was found by~\cite{Buote:2006kx}. The
halo parameters can thus be expressed in terms of $\rho_s$ and $r_s$
and are presented in Table~\ref{tab:Jbar}. This model is hereafter
referred as to RB02. A similar procedure was applied in the
\emph{Fermi}-LAT DM analysis of galaxy
clusters~\citep{2010JCAP...05..025A}.

A different approach is to use dynamical tracers of
the gravitational potential of the cluster halo, such as stars,
globular clusters or planetary nebulae. This method
is limited by the observability of such tracers, but can 
yield less model-dependent and more robust modeling of the DM
distribution. However, some uncertainty is introduced by the
translation of the tracer's velocity dispersion measurement into a
mass profile, which usually implies solving the Jeans equations under
some simplifying assumptions \citep{2008gady.book.....B}. From velocity dispersion measurements on dwarf galaxies observed up to about 1.4 Mpc, a dynamical analysis of the Fornax cluster by \citet{2001ApJ...548L.139D} constrained the cluster mass. The associated DM density profile, hereafter referred as to DW01, can be well described by a NFW profile~\citep{2008A&A...478L..23R} with parameters given in Table~\ref{tab:Jbar}.


 \cite{2008A&A...478L..23R} have analyzed the DM distribution in the inner regions of Fornax by using the globular clusters as dynamical tracers. This allowed an accurate DM mass profile measurement out to a
radial distance of 80 kpc from the galactic cluster centre, corresponding to
an angular distance of $\sim 0.25 \degr$. The resulting velocity dispersion measurements can be well fitted by a NFW DM halo profile with parameters given in Table~\ref{tab:Jbar}. This density profile (hereafter referred as to RS08) determination is in good agreement with the determination inferred from ROSAT-HRI X-ray measurements~\citep{Paolillo:2001kb}. Detailed analysis using subpopulations of globular clusters done in~\cite{2010AA...513A..52S} showed that both a NFW and a Burkert DM halo profiles can equally well fit the globular cluster velocity dispersion measurements. Representative DM halo profiles using different sets of globular clusters samples, hereafter referred as to SR10 a$_6$ and SR10 a$_{10}$, are extracted from Table 6 of~\cite{2010AA...513A..52S}. The parameters for both the NFW and Burkert DM halo profiles are given in Table~\ref{tab:Jbar}.


Using the dark matter halo parameters derived from the above-mentioned methods, values of \jbar were derived
for different angular integration radii. The point-spread-function of \hess corresponds to an
integration angle of $\sim 0.1^{\circ}$~\citep{2006A&A...457..899A},
and most often the smallest possible angle is used in the search for
dark matter signals in order to suppress  background events. However, since a
sizable contribution to the \gr flux may also arise from dark matter
subhalos located at larger radii (see Section \ref{sec:subs}),
integration angles of 0.5$^{\circ}$ and 1.0$^{\circ}$ were also considered.
The choice of the tracer samples induces a spread
in the values of the astrophysical factor \jbar up to one order of
magnitude for an integration angle of 0.1$^{\circ}$. Note that the
measurements of \cite{2008A&A...478L..23R} and
\cite{2010AA...513A..52S} trace the DM density distribution only up to
80 kpc from the center. In consequence the derived values of the virial mass and
radius are significantly smaller than those derived from  X-ray
measurements on larger distance scales~\citep[see for instance figure
  22 of][]{2010AA...513A..52S}. Thus the DM density values may be
underestimated for distances larger than about 100 kpc. On the other
hand, it is well known that for an NFW profile about 90\% of the DM
annihilation signal comes from the volume within the scale radius
$r_{\rm s}$. Therefore, even for NFW models with large virial radii
such as RB02 and DW01, the main contribution to the annihilation
signal comes from the region inside about 98 kpc and 220 kpc, respectively.

\begin{table}[h]
  \begin{center}
    \begin{tabular}{l|c|c|c c c }
      \hline \hline
            \multicolumn{3}{c}{}& \multicolumn{3}{|c}{\bf \jbardo [10$^{21}$ GeV$^2$cm$^{-5}$] }\\
         \hline
       &&  & \multicolumn{3}{|c}{\bf NFW profile}  \\
       Model & r$_s$ [kpc] & $\rho_s$ [$\rm M_{\odot} pc^{-3}$]  & $\theta_{\rm max}$ = 0.1\degr  & $\theta_{\rm max}$ = 0.5\degr & $\theta_{\rm max}$ = 1.0\degr\\
          \hline
       RB02 & 98 & 0.0058 & 112.0 & 6.5 & 1.7 \\
       DW01 & 220 & 0.0005 & 6.2 & 0.5 & 0.1 \\
       RS08 & 50 & 0.0065 & 24.0 & 1.2 & 0.3 \\
       SR10 a$_{10}$ & 34 & 0.0088 &15.0 & 0.6 & 0.1\\
       SR10 a$_6$ &200 & 0.00061 & 7.0 & 0.5 & 0.1  \\
       \hline
       &&& \multicolumn{3}{|c}{\bf Burkert profile}\\
       Model &r$_c$ [kpc]& $\rho_c$ [$\rm M_{\odot} pc^{-3}$] & $\theta_{\rm max}$ = 0.1\degr  & $\theta_{\rm max}$ = 0.5\degr & $\theta_{\rm max}$ = 1.0\degr\\
       \hline
       SR10 a$_{10}$& 12& 0.0728 & 15.0  &0.6 & 0.2  \\
       SR10 a$_6$ &94 &0.0031 & 2.4 & 0.5&  0.1 \\
       \hline \hline
    \end{tabular}
    \caption{Dark matter halo models for the Fornax galaxy cluster. The first three
      columns show the selected profiles discussed in Section~\ref{sec:fornaxdm} with their respective
      NFW or Burkert halo parameters. The last three columns show the
      astrophysical factor $\jbar$, calculated for three different integration radii. }
      \label{tab:Jbar}
  \end{center}
\end{table}

\subsection{Dark matter halo substructures}
\label{sec:subs}
Recent cosmological N-body simulations, such as Aquarius~\citep{2008MNRAS.391.1685S}
and Via Lactea~\citep{2008Natur.454..735D}, have suggested the presence
of dark matter substructures in the form of self-bound overdensities
within the main halo of galaxies. A quantification of the substructure
flux contribution to the total \gr flux was computed from the Aquarius simulation by
\cite{2009PhRvL.103r1302P} using the NFW profile RB02 as the DM density distribution of the smooth halo\footnote{This halo is also well suited with respect to the others discussed in Section\ref{sec:fornaxdm} since substructures in the form of gravitationally bound dwarf galaxies to Fornax are observed up to about 1 Mpc. They are thus included within the virial radius predicted by the RB02 profile ($R_{\rm vir}$ $\simeq$ 1 Mpc).}. The substructure enhancement over the smooth host halo contribution along the line of sight is defined as $B_{\rm sub}(\Delta \Omega) = 1 + \LL_{\rm sub}(\Delta
\Omega)/\LL_{\rm sm}(\Delta \Omega)$, where $\LL_{\rm sm/sub}(\Delta
\Omega)$ denotes the annihilation luminosity of the smooth host halo
and the additional contribution from substructures, respectively. The
former is defined by:
\begin{equation}
  \LL_{\rm sm/sub}(\Delta \Omega) \,=\, \Delta\Omega\times\overline{J}_{\rm sm/sub}(\Delta \Omega)\,=\, \int_{\Delta\Omega} \dd \Omega
  \int_{\textrm{l.o.s.}} \dd l \times \rho^2_{\rm sm/sub}[r(l)] \, ,
\label{eq:Lumsm}
\end{equation}
  where $\rho_{\rm sm/sub}$ is the DM density distribution of the smooth halo and substructures, respectively. In order to perform the LOS integration over the subhalo contribution, an effective substructure density
$\tilde{\rho}_{\rm sub}$ is parametrized following \cite{2008MNRAS.391.1685S} and
\cite{2009PhRvL.103r1302P} as:
\begin{equation}
\tilde{\rho}_{\rm sub}^2(r) = \frac{A(r)\,0.8 C \, \LL_{\rm sm}(R_{\rm vir})}{4\pi r^2 R_{\rm vir}}\left(\frac{r}{R_{\rm vir}}\right)^{-B(r)} \, ,
\label{eq:Lumsub}
\end{equation}
where
\begin{equation}
A(r) = 0.8 - 0.252\,\textrm{ln}(r/R_{\rm vir})
\end{equation}
and
\begin{equation}
B(r) = 1.315 - 0.8(r/R_{\rm vir})^{-0.315} \, .
\end{equation}
$\rm \LL_{sm}(R_{\rm vir})$ is the smooth halo luminosity within the
virial radius $R_{\rm vir}$. The normalization is given by $C =
(M_{\rm min}/M_{\rm lim})^{0.226}$, where $M_{\rm min} = 10^5
M_{\odot}$ is the minimum substructure mass resolved in the simulation
and $M_{\rm lim}$ is the intrinsic limiting mass of substructures, or
free-streaming mass.  A conventional value for this quantity is
$M_{\rm lim} = 10^{-6} M_{\odot}$~\citep{2006ApJ...649....1D},
although a rather broad range of values, down to $M_{\rm lim} = 10^{-12} M_{\odot}$, is possible for different
models of particle dark matter~\citep{2009NJPh...11j5027B}.  Assuming
a specific DM model, a constraint on $M_{\rm lim}$ was derived by
\citet{2009PhRvL.103r1302P} using EGRET \gr upper limits on the Virgo
cluster and a lower bound was placed at \mbox{$M_{\rm lim} = 5\times
10^{-3} M_{\odot}$.} Nevertheless, the effect of a smaller limiting
mass  is also investigated in this work.

\begin{figure}[ht]
  \centering
  \includegraphics[width=0.7\textwidth]{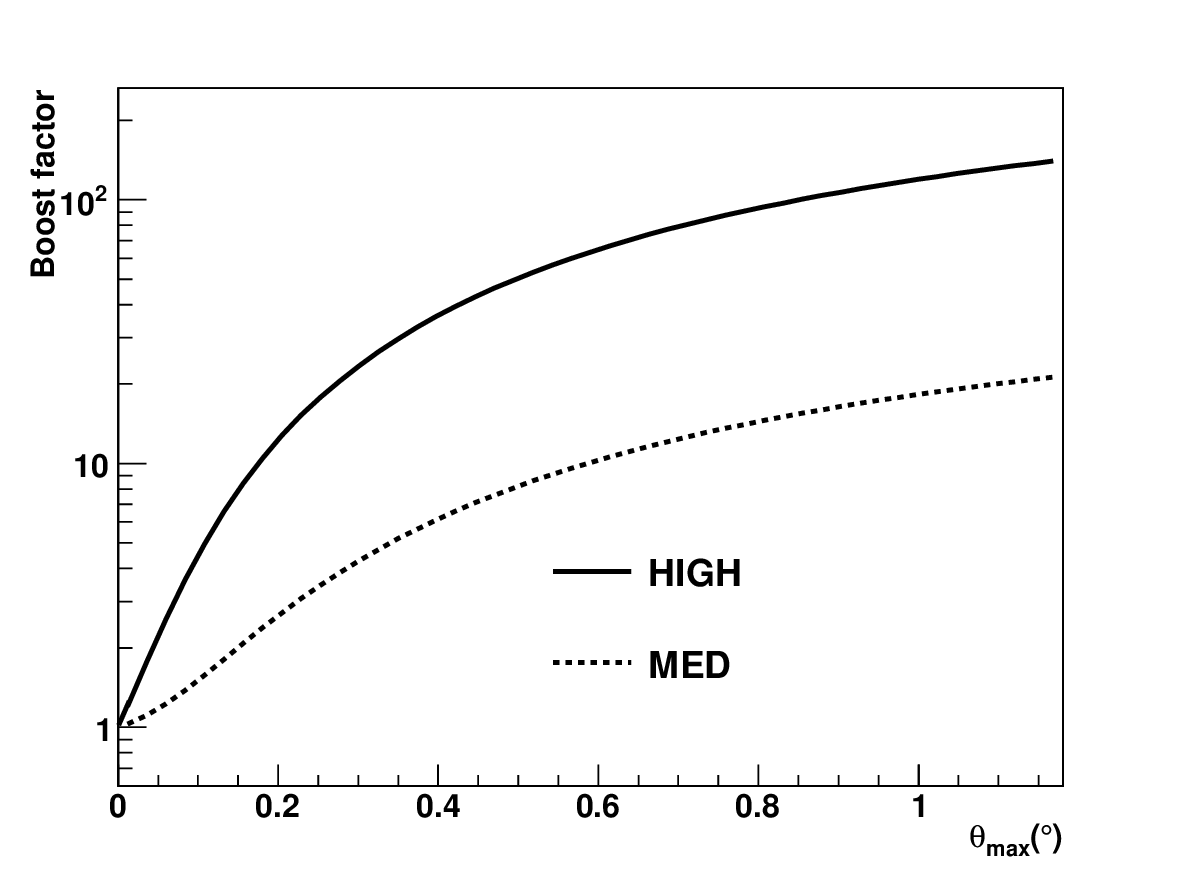}
  \caption{Substructure \gr flux enhancement as function of the
    opening angle of integration. Two values of the limiting mass of
    substructures are used: $M_{\rm lim} = 10^{-6}M_{\odot}$, for the
    high (HIGH) boost (solid line), and $M_{\rm lim} = 5\times 10^{-3}
    M_{\odot}$, for the medium (MED) boost (dashed line). The RB02
    profile is chosen as the smooth host DM halo.}
  \label{fig:luminosities}
\end{figure}

Figure~\ref{fig:luminosities} shows the substructure enhancement
$B_{\rm sub}$ over the smooth halo as function of the opening
integration angle. At the distance of Fornax, integration regions
larger than $\sim$ 0.2$^{\circ}$ correspond to more than 65
kpc. Beyond these distances the substructure enhancement exceeds a
factor 10. This justifies {\it extended} analyses using integration
angles of 0.5$^{\circ}$ and
1.0$^{\circ}$. Two values of the limiting mass of substructures are
used: $M_{\rm lim} = 10^{-6}M_{\odot}$ and $M_{\rm lim} = 5\times
10^{-3} M_{\odot}$, inducing a \emph{high} and a \emph{medium} value
of the enhancement, respectively. The values of $B_{\rm sub}$ for the opening
angles of 0.1$^{\circ}$, 0.5$^{\circ}$ and 1.0$^{\circ}$ and for both
values of $M_{\rm lim}$ are given in Table~\ref{tab:boost}. These
values are larger than those derived in~\cite{2010JCAP...05..025A}. In
their study the substructure enhancement is calculated from the Via
Lactea~\citep{2008Natur.454..735D} simulation, where a different
concentration mass relation is obtained. For a careful comparison see
\citet{2011PhRvD..83b3518P}. 

\begin{table}[bht]
\centering
\begin{tabular}{ c || c  c  c }
 \hline  \hline
 $\theta_{\rm max} $ & 0.1$^\circ$ & 0.5$^\circ$ & 1.0$^\circ$\\
 \hline \hline
  $M_{\rm lim} = 10^{-6} M_{\odot}$ &  4.5  & 50.5 & 120  \\
  \hline
  $M_{\rm lim} = 5 \times 10^{-3} M_{\odot}$ &  1.5  & 8.2 & 18.3  \\
  \end{tabular}
\caption{Enhancement $B_{\rm sub}$ due to the halo substructure
  contribution to the DM flux, for different opening angles of
  integration $\theta_{\rm max} $. The enhancement is calculated for
  two limiting masses of substructures $M_{\rm lim}$ and over the smooth DM halo RB02.}
\label{tab:boost}
\end{table}

\section{Observations and data analysis}
\label{sec:hess}
The High Energy Stereoscopic System (H.E.S.S.) consists of four
identical imaging atmospheric Cherenkov telescopes. They are located
in the Khomas Highland of Namibia ($23^{\circ} 16^{\prime} 18^{\prime
  \prime}$ South, $16^{\circ}30^{\prime}00^{\prime \prime}$ East) at
an altitude of 1800~m above sea level. The H.E.S.S. array was designed
to observe VHE \grs through the Cherenkov light emitted by
charged particles in the electromagnetic showers initiated by these
\grs when entering the atmosphere. Each telescope has an
optical reflector consisting of 382 round facets of 60~cm diameter
each, yielding a total mirror area of
107~m$^2$~\citep{2003APh....20..111B}. The Cherenkov light is
focused on cameras equipped with 960 photomultiplier tubes, each one
subtending a field of view of $0.16^\circ$. The total field of view
is $\sim$$5^\circ$ in diameter. A stereoscopic reconstruction of the shower is
applied to retrieve the direction and the energy of the primary $\gr$.

Dedicated observations of the Fornax cluster, centered on NGC 1399, were conducted in fall 2005~\citep{2008ICRC....3..933P}. They were carried out in \emph{wobble mode}~\citep{2006A&A...457..899A}, \textit{i.e.} with the target typically
offset by $0.7^\circ$ from the pointing direction,
allowing simultaneous background estimation from the same
field of view. The total data passing the standard \hess data-quality
selection~\citep{2006A&A...457..899A} yield an exposure of 14.5 hrs
live time with a mean zenith angle of 21$^\circ$.

The data analysis was performed using an improved \textit{model}
analysis as described in~\cite{2009APh....32..231D}, with independent
cross-checks performed with the \textit{Hillas}-type analysis
procedure~\citep{2006A&A...457..899A}. Both analyses give compatible results.
 Three different signal integration angles were used, $0.1^\circ$,
$0.5^\circ$ and $1^\circ$. The cosmic-ray background was estimated
with the {\it template} model~\citep{2003A&A...410..389R}, employing the
source region, but selecting only hadron-like events from image cut
parameters.

\begin{figure}[hb]
  \centering
  \includegraphics[width=\textwidth]{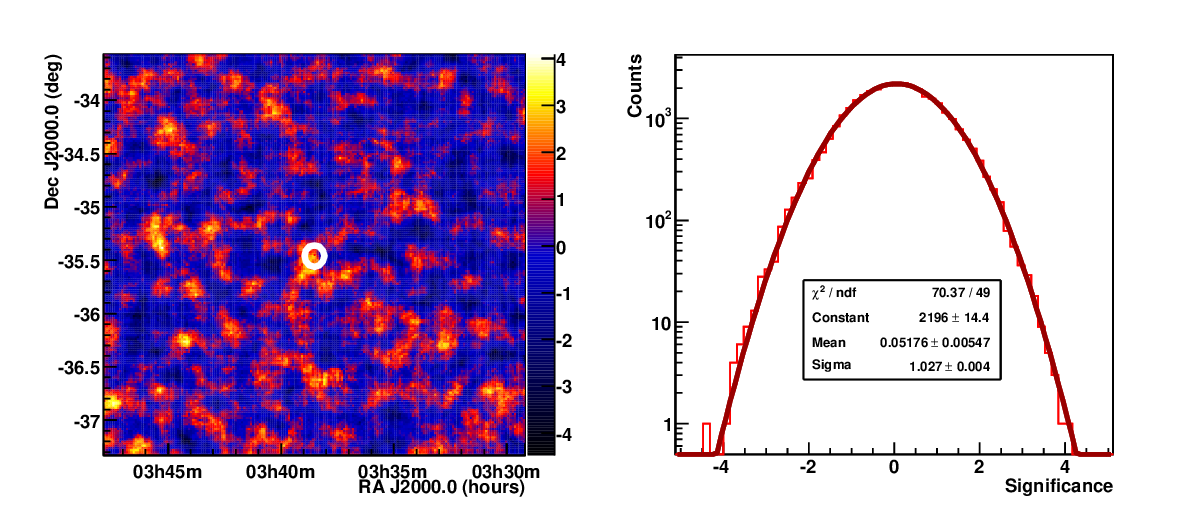}
  \caption{Left: Significance map in equatorial
    coordinates, calculated according to the Li $\&$ Ma
    method~\protect\citep{1983ApJ...272..317L}, with an oversampling radius of
    $0.1^\circ$ . The white circle denotes the
    $0.1^\circ$ integration region. No significant excess is seen at
    the target position. Right: Distribution of the
    significance. The solid line is a Gaussian fitted to the data. The
    significance distribution is well described by a normal
    distribution.
    \label{fig:significance}}
  \end{figure}

No significant excess was found above the background level in any of
the integration regions, as visible in Fig.~\ref{fig:significance} for
an integration angle of $0.1^{\circ}$. An
upper limit on the total number of
observed $\grs$, $N_{\gamma}^{95\%\,{\rm C.L.}}$, was calculated
at 95$\%$  confidence level (C.L.).  The calculation followed the
method described in \cite{1998PhRvD..57.3873F}, using the number of
\gr candidate events in the signal region $N_{\rm ON}$ and the
\emph{normalized} number of \gr events in the background region
$\overline{N}_{\rm OFF}$. Since the normalization is performed with respect to the direction-dependent
acceptance and event rate, the background normalization
factor for $\overline{N}_{ \rm OFF}$ as defined in
\cite{2003A&A...410..389R} is $\alpha \equiv 1$.
This is equivalent to the assumption that the uncertainty on the background
determination is the same as for the signal, allowing a conservative
estimate of the upper limits. This information is summarized in
Table~\ref{tab:ngammalimits}.

A minimal \gr energy ($E_{\rm min}$) is defined as the energy at
which the acceptance for point-like observations reaches 20\% of its
maximum value, which gives 260 GeV for the observations of Fornax. Limits on the number
of \gr events above the minimal energy $E_{\rm min}$ have also been
computed (see Table~\ref{tab:fluxlimits}) and are used
in Section~\ref{sec:fluxlimits} for the calculation of upper limits on
the \gr flux.

\begin{table}[bht]
\centering
\begin{tabular}{ c | c | c | c | c}
  $\theta_{\rm max}$ & N$_{\rm ON}$ & $\overline{\rm N}_{\rm OFF}$ &
  $N_{\gamma}^{95\%\,{\rm C.L.}}$ & Significance \\
\hline
0.1$^{\circ}$ &  160 & 122 & 71 & 2.3 \\
0.5$^{\circ}$ &  3062 & 2971 & 243 &  1.2 \\
1.0$^{\circ}$ &  11677 & 11588 & 388 & 0.6 \\
\hline 
\end{tabular}
\caption{Numbers of VHE \gr events from the
  direction of the Fornax galaxy cluster centre, using three different opening angles for the
  observation. Column 1 gives the opening angle $\theta_{\rm max}$,
  columns 2 and 3 the numbers of \gr candidates in the ON region,
  N$_{\rm ON}$, and the normalized number of \gr in the OFF region,
  $\rm \overline{N}_{OFF}$,  respectively. Column 4 gives the
  $95\%\,{\rm C.L.}$ upper  limit on the number of \gr events
  according to  \protect\cite{1998PhRvD..57.3873F}. The significance of the numbers of \gr candidates in the ON region is stated in column 5 according to~\protect\cite{1983ApJ...272..317L}.}
\label{tab:ngammalimits}
\end{table}

\section{\gr flux upper limits}
\label{sec:fluxlimits}

Upper limits on the number of observed $\gr$s above a minimal energy
$E_{\rm min}$ can be translated into an upper limit on the observed
\gr flux $\Phi_\gamma$ if the energy spectrum $\dd N_{\gamma} /\dd E_{\gamma} $ of the source is
assumed to be known, as indicated by equation \ref{eq:fluxlimit}.

\begin{equation} \Phi_\gamma^{\rm 95\%\,C.L.}(E_{\gamma}>\Eth) =
\frac{N_{\gamma}^{95\%\,{\rm C.L.}}(E_{\gamma}>\Eth) \displaystyle\int_{\Eth}^{\infty}\dd E_{\gamma} \frac{\dd N_{\gamma}}{\dd E_{\gamma}}(E_{\gamma})}{T_{\rm
{obs}} \displaystyle\int_{\Eth}^{\infty}\dd E_{\gamma} \, A_{\rm
{eff}}(E_{\gamma}) \frac{\dd N_{\gamma}}{\dd E_{\gamma}}(E_{\gamma})}
\, .
\label{eq:fluxlimit}
\end{equation}

\noindent
Here, $T_{\rm obs}$ and $A_{\rm eff}$ denote the target observation time and
the instrument's effective collection area, respectively. The intrinsic spectra of standard astrophysical VHE $\gr$ sources
\citep{2009ARA&A..47..523H} typically follow power-law behavior of index
$\Gamma \approx 2-3$. Upper limits at 95\% C.L. on the integral flux above the minimum energy
(cf. Section~\ref{sec:hess}) are given in Table~\ref{tab:fluxlimits}
for different source spectrum indices.

\begin{figure}[hb]
  \centering
  \includegraphics[width=\textwidth]{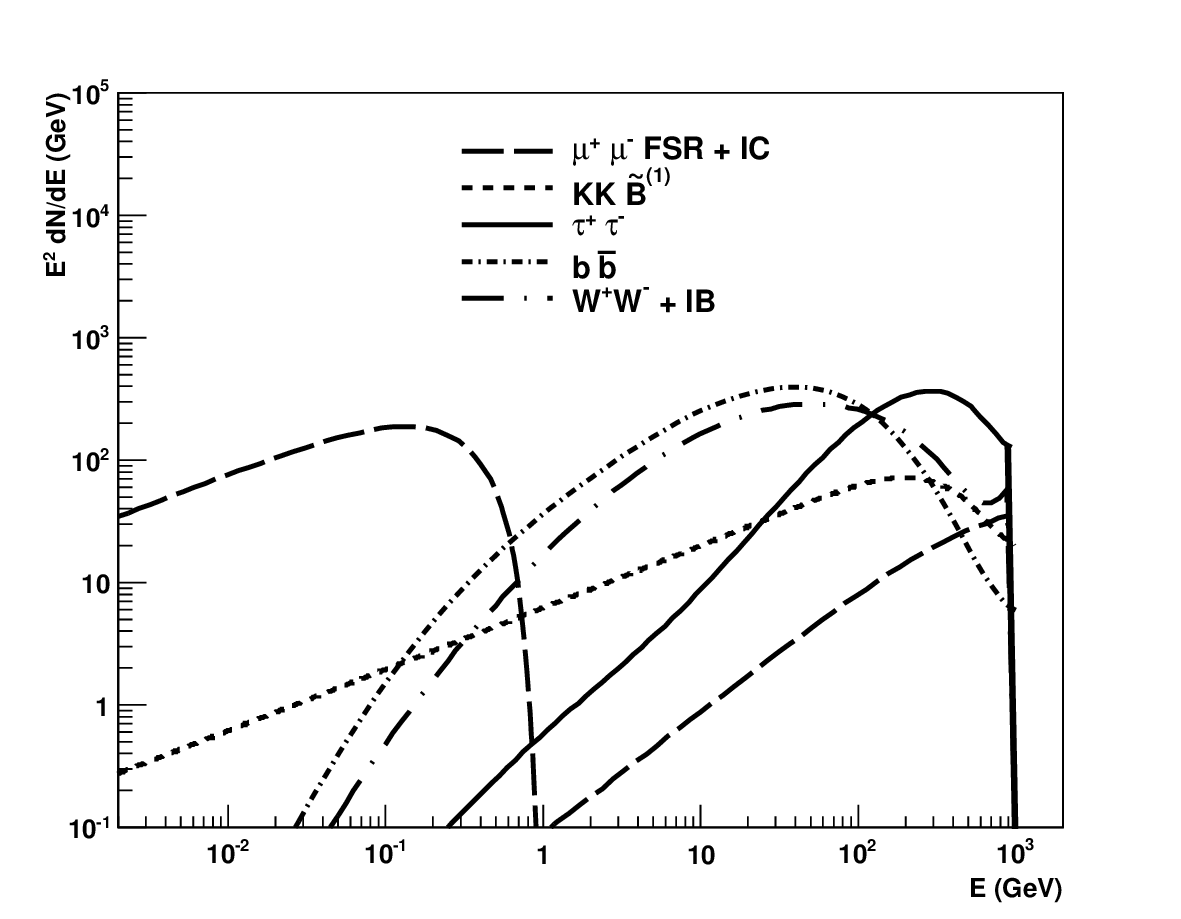}
  \caption{Photon spectra for 1 TeV dark matter particles
    self-annihilating in different channels. Spectra from DM annihilating purely into $b\overline{b}$ (dot-dashed line), $\tau^+\tau^-$ (black solid line) and $W^+W^-$ (long-dashed dotted line) are shown.
    The latter shows the effect of Internal Bremsstrahlung (IB) occuring for the $W^+W^-$ channel. The \gr spectrum from the annihilation of $\widetilde{B}^{(1)}$ hypergauge boson pairs arising in Kaluza-Klein (KK) models with UED is also plotted (dotted line). The long dashed line show the photon spectra from final-state radiation (FSR) and the inverse Compton (IC) scattering contribution in the case of DM particles annihilating into muon pairs.}
  \label{fig:spectra}
\end{figure}

\begin{figure}[hb]
  \centering
  \includegraphics[width=\textwidth]{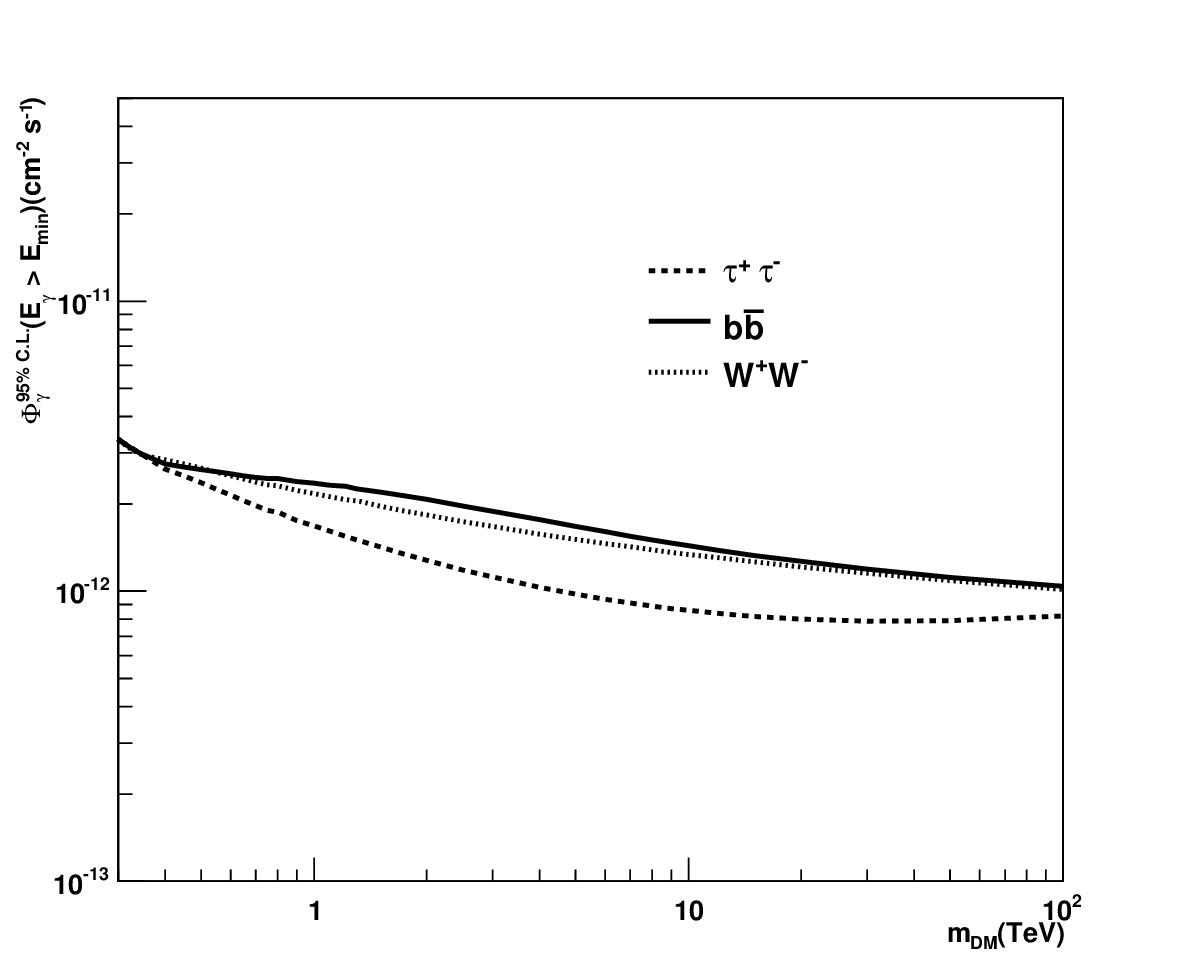}
  \caption{Upper limits $95\%$ C.L. on the \gr flux as a function of the DM particle mass for E$_{\rm min}$= 260 GeV from the direction of Fornax. DM particles annihilating into $b\bar{b}$ (solid line) , $W^+W^-$ (dotted line) and $\tau^+\tau^-$ (dashed line) pairs are considered.}
  \label{fig:flux95DM}
\end{figure}

Dark matter annihilation spectra depends on the assumed
annihilation final states of the DM model. For instance, some supersymmetric extensions of the Standard Model~\citep{1996PhR...267..195J} predict the \emph{neutralino} as the lightest stable supersymetric particle, which
would be a good dark matter candidate. In general, the
self-annihilation of neutralinos will give rise to a continuous \gr
spectrum from the decay of neutral pions, which are produced in the
hadronisation process of final-state quarks and gauge bosons. Universal extra-dimensional (UED) extensions of
the SM also provide suitable DM candidates. In these models, the first
Kaluza-Klein (KK) mode of the hypercharge gauge boson
$\widetilde{B}^{(1)}$ is the lightest KK particle (LKP) and it can be a DM particle candidate~\citep{Servant:2002aq}. Nevertheless, in the absence of a preferred DM particle model, constraints are presented here in a model-independent way, i.e. for given pure pair annihilation final state for the DM pair annihilation processes and DM particle mass. The only specific DM particle model studied here is the KK $\widetilde{B}^{(1)}$ particle model, where the branching ratios of each annihilation channel are known. A wide range of dark matter masses is investigated from about 100 GeV up to 100 TeV. A model-independent upper bound on the dark matter mass can be derived from unitarity for thermally produced DM as done in the seminal paper of \citet{1990PhRvL..64..615G} and subsequent studies of \citet{2007PhRvL..99w1301B} and \cite{2008PhRvD..78f3542M}. Assuming the current DM relic density measured by WMAP~\citep{2011ApJS..192...16L}, the inferred value is about 100 TeV. Figure~\ref{fig:spectra} shows different annihilation spectra for 1 TeV mass dark matter particles. Spectra of DM particles annihilating into $b\bar{b}$, $W^+W^-$ and $\tau^+\tau^-$  pairs are extracted from \cite{Cirelli:2010xx}, and calculated from \cite{Servant:2002aq} for Kaluza-Klein $\widetilde{B}^{(1)}$ annihilation. Flux upper limits as function of the DM particle mass are presented in Figure~\ref{fig:flux95DM} assuming DM annihilation purely into $b\bar{b}$ , $W^+W^-$ and $\tau^+\tau^-$ and an opening angle of the integration of 0.1\grad. Flux upper limits reaches 10$^{-12}$ cm$^{-2}$ s$^{-1}$ for 1 TeV DM mass.

\begin{table}[bht]
\centering
\begin{tabular}{ c | c | c  c }
   \multirow{2}{*}{$\theta_{\rm max}$ } & \multirow{2}{*}{$N_{\gamma}^{95\%\,{\rm C.L.}}(E_{\gamma} >
   E_{\rm min})$ } & \multicolumn{2}{c}{$\Phi_\gamma^{\rm 95\%\,C.L.}
     (E_{\gamma}>\Eth) (10^{-12}$ cm$^{-2}$ s$^{-1})$}   \\
 &  & $\qquad\Gamma = 1.5\qquad$  & $\Gamma = 2.5 $ \\
\hline
0.1\grad & 41.3 & 0.8 & 1.0  \\
0.5\degr & 135.1&  2.3 & 3.3 \\
1.0\grad & 403.5&  6.8 & 10.0  \\
\end{tabular}
\caption{Upper limits on the VHE \gr flux from the direction of Fornax, assuming a power-law spectrum with spectral index $\Gamma$ between 1.5 and 2.5. Column 1 gives the opening angle of the
  integration region $\theta_{\rm max}$, column 2 the upper limits on
  the number of observed \grs above   the minimum energy $E_{\rm min}
  = 260 \gev$, calculated at 95$\%$ C.L.. Columns 3 and 4 list the 95$\%$
  C.L. integrated flux limits above the minimum energy, for two power
  law indices.}
\label{tab:fluxlimits}
\end{table}

Recent studies~\citep{2009PhRvD..80b3005J,2009PhRvL.103r1302P,2010MNRAS.409..449P}
have computed the cosmic-ray induced \gr flux from pion decays using a
cosmological simulation of a sample of 14 galaxy
clusters~\citep{2008MNRAS.385.1211P}. Since the electron induced \gr flux from inverse Compton is
found to be systematically subdominant compared to the pion decay \gr flux~\citep{2009PhRvD..80b3005J}, this contribution is not considered. Using the results of \cite{2009PhRvL.103r1302P},  the \gr flux above 260 GeV for
Fornax is expected to lie between a few 10$^{-15}$ cm$^{-2}$ s$^{-1}$
and 10$^{-14}$ cm$^{-2}$ s$^{-1}$ for an
opening angle of observation of 1.0$^{\circ}$. The flux is about 2-to-3 orders of magnitude lower than the upper limits presented in Table~\ref{tab:fluxlimits}, thus this scenario cannot be constrained.

Assuming a typical value of the annihilation cross-section for thermally-produced DM, $\langle\sigma v\rangle\, = 3\times$10$^{-26}$ cm$^3$s$^{-1}$, a mass of 1 TeV and the NFW profile of DM density profile of Fornax RB02, the predicted DM \gr flux is found to be  a few  10$^{-13}$ cm$^{-2}$ s$^{-1}$. This estimate takes into account the \gr enhancement due to dark halo substructure and the Sommerfeld enhancement (see section~\ref{sec:exclusion}) to the overall DM \gr flux. Therefore the
dominant \gr signal is expected to originate from DM annihilations. Constraints on the DM-only scenario are derived in the
following section.

\section{Exclusion limits on dark matter annihilations}
\label{sec:exclusion}
\begin{figure}[hb]
  \centering
  \includegraphics[width=\textwidth]{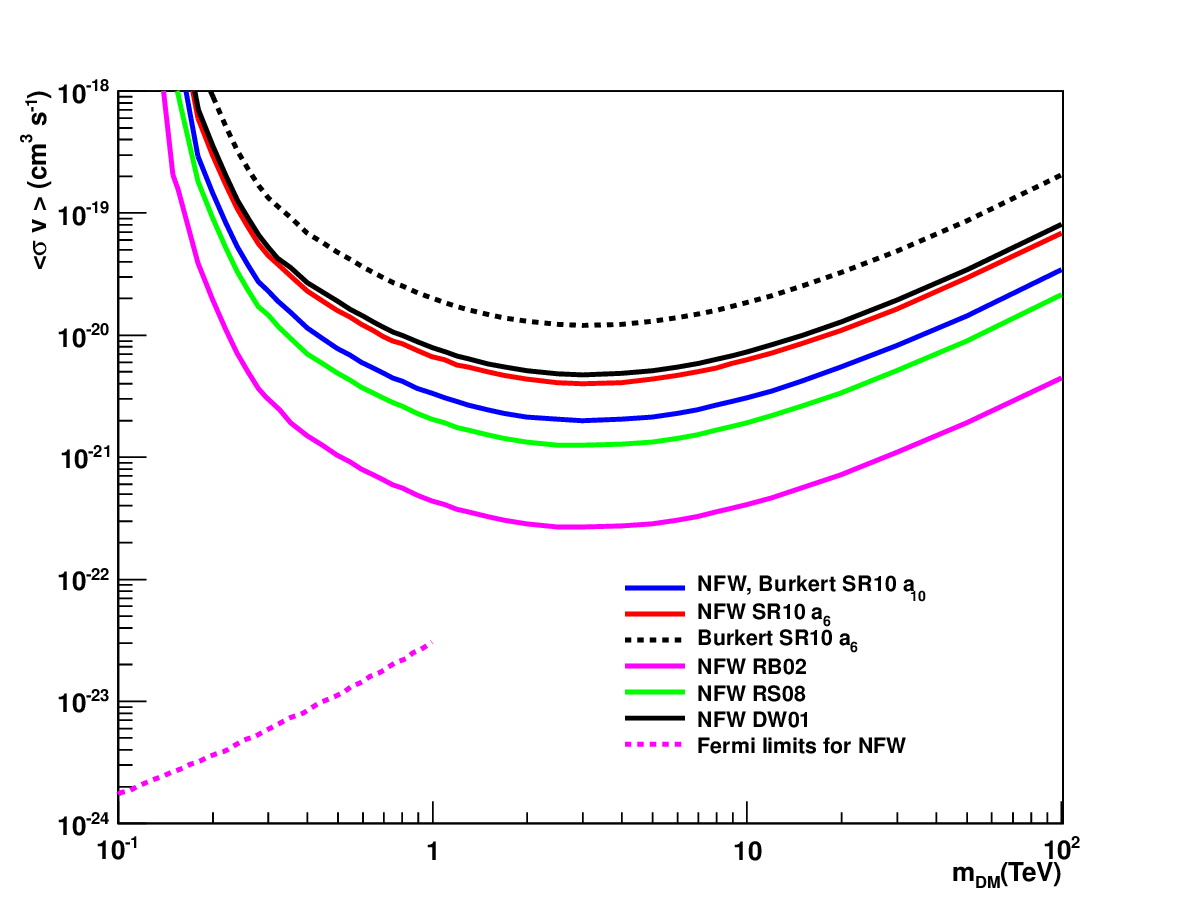}
  \caption{Upper limit at $95\%$ C.L. on the velocity-weighted
    annihilation cross-section \sigmav
    as a function of the DM particle mass, considering DM particles
    annihilating purely into $b\overline{b}$ pairs. The limits are given for an integration angle
    $\theta_{\rm max}$ = 0.1\degr. Various DM halo profiles are considered:  NFW profiles, SR10 a$_{10}$ (blue solid line), DW01 (black solid line), RB02 (pink solid line) and RS08 (green solid line), and Burkert profiles, SR10 a$_6$ (red dotted line) and a$_{10}$ (blue solid line). See Table \ref{tab:Jbar} for more details. The \emph{Fermi}-LAT upper
    limits~\protect\citep{2010JCAP...05..025A} for the NFW profile RB02 are also plotted.}
  \label{fig:exclusionsAllHaloes}
\end{figure}

\begin{figure}[hb]
  \centering
  \includegraphics[width=\textwidth]{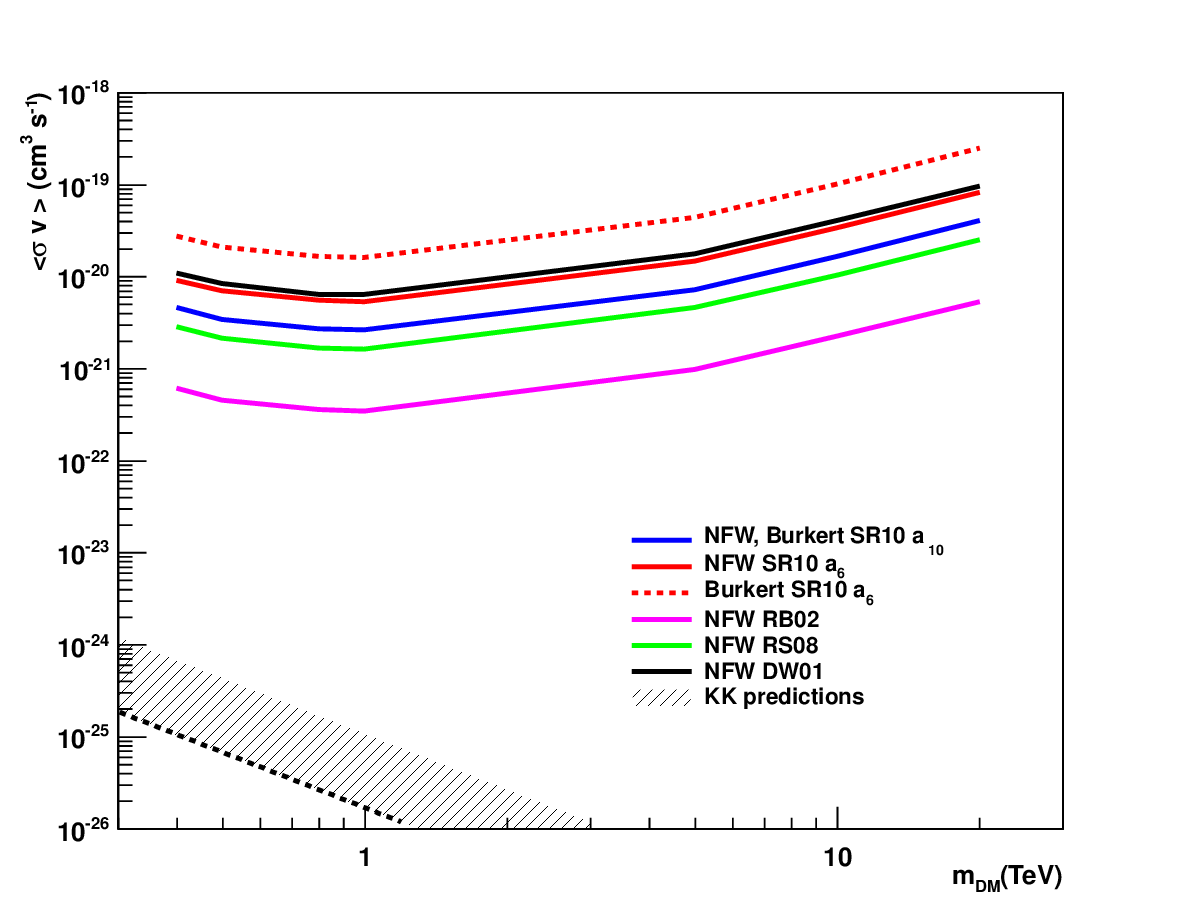}
  \caption{Kaluza-Klein hypergauge boson $\tilde{B}^{(1)}$ dark matter:  Upper limit at $95\%$ C.L. on $\left\langle \sigma v
    \right\rangle$ as function of the $\tilde{B}^{(1)}$ mass towards Fornax. The limits are given for an integration angle $\theta_{\rm max}$ = 0.1\degr.
    The NFW profiles, SR10 a$_{10}$ (blue solid line), DW01 (black solid line), RB02 (pink solid line) and RS08 (green solid line), and Burkert profiles, SR10 a$_6$ (red dotted line) and a$_{10}$ (blue solid line). See Table \ref{tab:Jbar} for more details. The prediction of $\left\langle \sigma v \right\rangle$ as function of the $\tilde{B}^{(1)}$ mass is given (dotted-line). A range for this predictions is given in case of a mass splitting between the LKP and the next LKP down to 1$\%$ (dashed area). }
  \label{fig:exclusionsKK}
\end{figure}

\begin{figure}[hb]
  \centering
  \includegraphics[width=\textwidth]{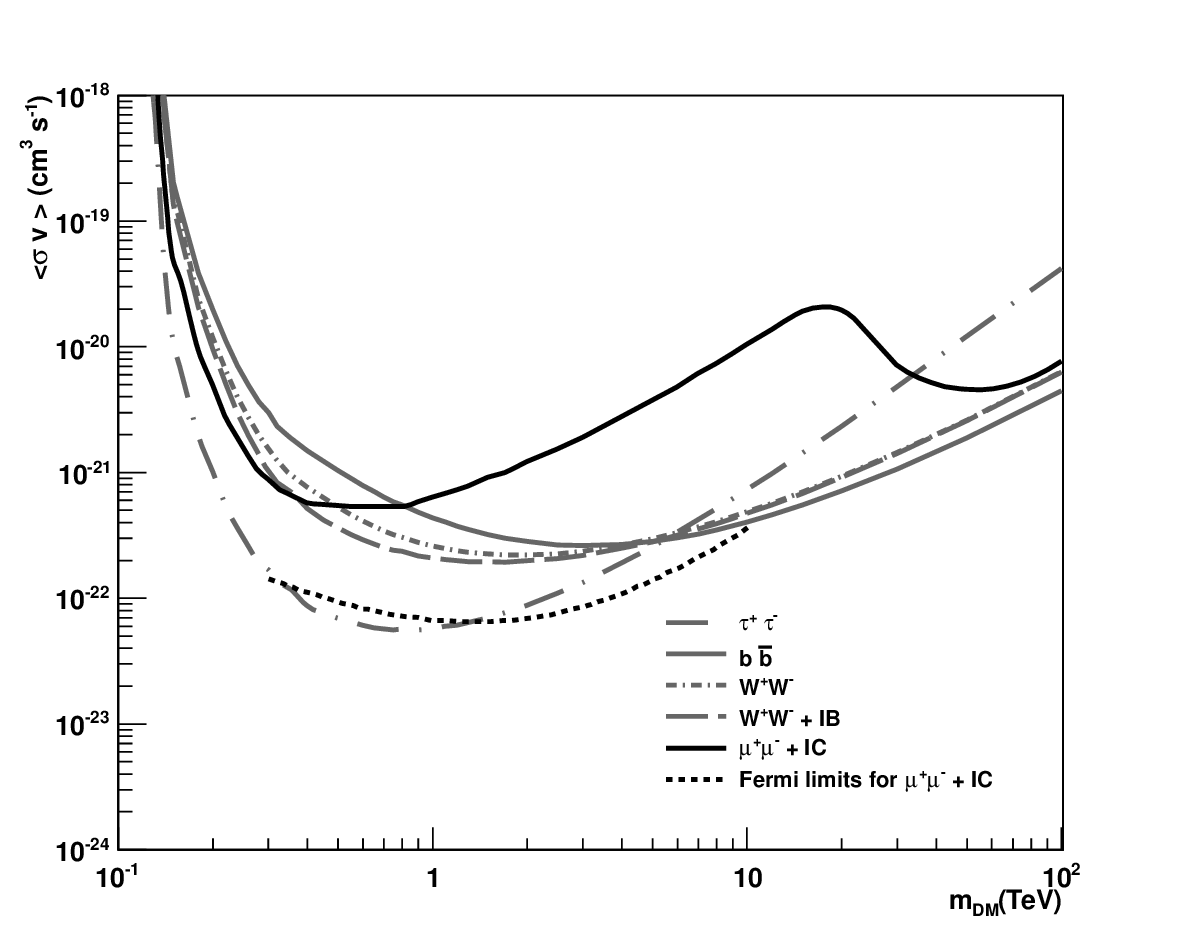}
  \caption{The effect of different DM particle models: Upper limit at $95\%$ C.L. on $\left\langle \sigma v
    \right\rangle $ as function of the DM particle mass.
    The limits are given for $\theta_{\rm max}$ = 0.1\degr and the NFW profile RB02.
    The limits  are shown for DM particles annihilating into $b\bar{b}$ (gray solid line) , $W^+W^-$ (gray dash-dotted line), $\tau^+\tau^-$ (gray long-dash-dotted line) pairs. The effect of Internal Bremsstrahlung (IB) occuring for the $W^+W^-$ channel is plotted in gray long-dashed line. The black solid line shows the limits for DM annihilating into $\mu^+\mu^-$ pairs including the effect of inverse Compton (IC) scattering. The \emph{Fermi}-LAT upper limits~\protect\citep{2010JCAP...05..025A} for the NFW profile RB02 and for an DM annihilating into $\mu^+\mu^-$ pairs including the effect of IC scattering are also plotted (black dotted line). See section \ref{sec:subs} for more details. }
  \label{fig:exclusionsAll}
\end{figure}

Upper limits at $95\%$ C.L. on the dark matter velocity-weighted annihilation cross-section can be derived
from the following formula:
\begin{equation}
\langle \sigma v \rangle^{\rm  95\%\,C.L.} =
\frac{8\pi}{T_{\rm {obs}}}\frac{m^2_{\rm DM}}{\jbar(\Delta
\Omega)\Delta\Omega}\,\frac{N_{\gamma}^{\rm  95\%\,C.L.}}{\displaystyle
\int_{0}^{m_{\rm DM}}\dd E_{\gamma} \, A_{\rm {eff}}(E_{\gamma})
\frac{\dd N_{\gamma}(E_{\gamma})}{\dd E_{\gamma}}} \, .
\label{eq:sigmavlimit}
\end{equation}
The factor $\jbar$ is extracted from Section~\ref{sec:jbar}. 
The exclusion limits as a function of the DM particle mass $m_{\rm DM}$ for different DM halo profile models are depicted in Figures
\ref{fig:exclusionsAllHaloes} and \ref{fig:exclusionsKK} for DM particles annihilating  exclusively into $b\overline{b}$ and $\widetilde{B}^{(1)}$ particles, respectively. Predictions for $\left\langle \sigma v \right\rangle$ as function of the $\widetilde{B}^{(1)}$ particle mass are given in Figure~\ref{fig:exclusionsKK} within the UED framework of \citet{Servant:2002aq}. As an illustration of a possible change in this prediction, a range of predicted $\left\langle \sigma v \right\rangle$ is extracted from Figure~2 of ~\citet{2008PhRvD..78e6002A}, in the case of a mass splitting between the LKP and the next lightest KK particle down to 1$\%$.
In the TeV range the $95\%$ C.L. upper limit on the annihilation cross-section $\left\langle \sigma v \right\rangle$ reaches $10^{-22}$~cm$^{3}$s$^{-1}$. Exclusion limits as a function of the DM particle mass $m_{\rm DM}$ assuming DM particle annihilating into $b\overline{b}$, $\tau^+\tau^-$ and $W^+W^-$ are presented in Figure~\ref{fig:exclusionsAll} for the RB02 NFW profile. Stronger constraints are obtained for masses below 1 TeV in the $\tau^+\tau^-$ where the $95\%$ C.L. upper limit on $\left\langle \sigma v \right\rangle$ reaches $10^{-23}$~cm$^{3}$s$^{-1}$.
The \emph{Fermi}-LAT exclusion limit for Fornax
is added in Figure \ref{fig:exclusionsAllHaloes} (pink dashed-line),
extending up to 1~TeV~\citep{2010JCAP...05..025A}. It is
based on the RB02 NFW profile and a $\gr$
spectrum which assumes annihilation to $b\overline{b}$ pairs.
Below  1~TeV, the \emph{Fermi}-LAT results provide
stronger limits than the H.E.S.S. results. However, the H.E.S.S. limits well complement the DM constraints in the TeV range.

Other DM particle models give rise to modifications of the $\gr$ annihilation spectrum which may increase the
predicted $\gr$ flux. Some of them are considered in the following.

\subsection{Radiative correction: Internal bremsstrahlung}
\label{sec:dmmodels}

\begin{figure}[hb]
  \centering
  \includegraphics[width=\textwidth]{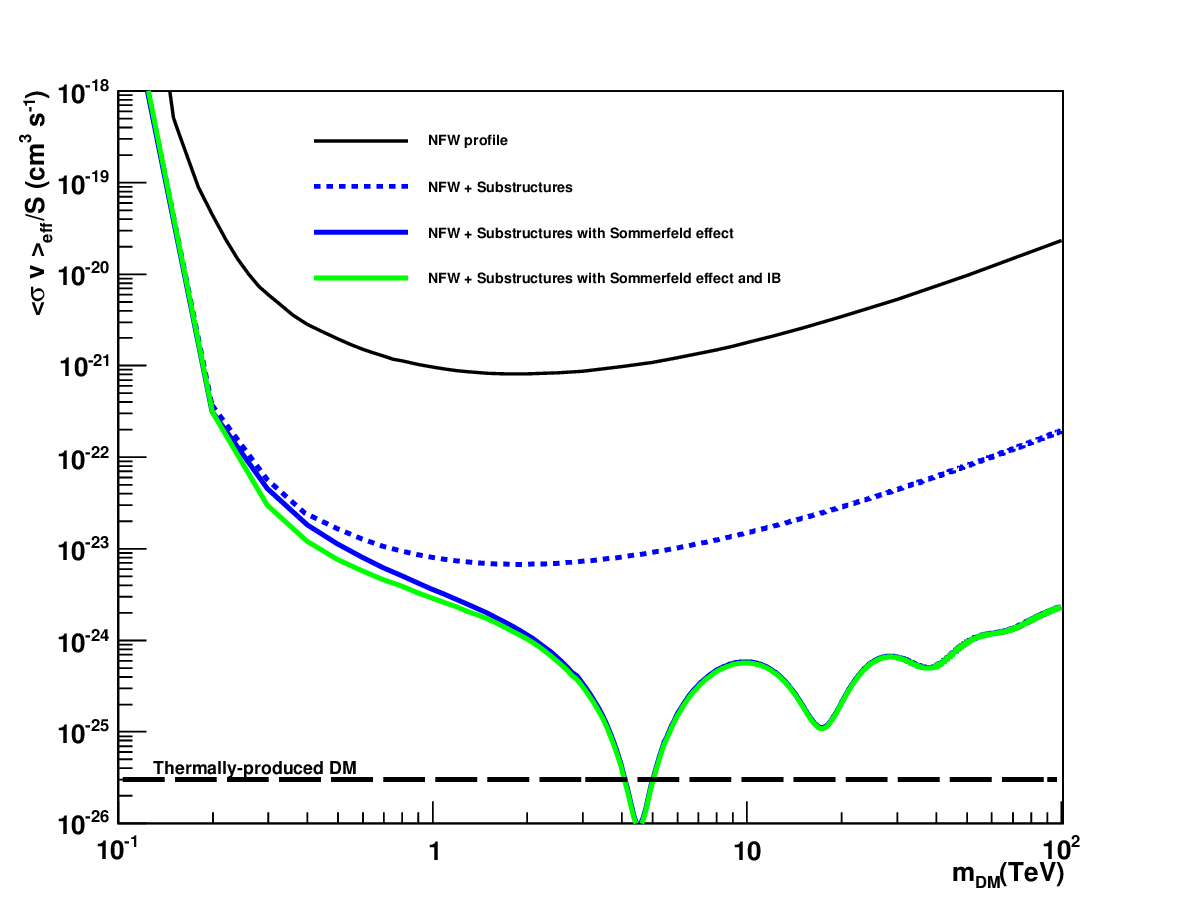}
  \caption{The Sommerfeld effect:
    Upper limits at $95\%$ C.L. on the effective annihilation
    cross-section $\sigmav_{\rm eff} = \sigmav_{0}/S$  as a function of
    the DM particle mass annihilating into $W$ pairs. The black line denotes the cross-section
    limit for $\theta_{\rm max} = 1.0\degr$ without \gr flux
    enhancement, the dashed blue line shows the effect
    of halo substructure (using the ``high boost'', cf. Fig.~\ref{fig:exclusions4}). The solid green and blue lines show the
    limit for the case of Wino dark matter annihilation enhanced by
    the Sommerfeld effect, with and without including Internal
    Bremsstrahlung, respectively.  The DM halo model RB02  is
    used (see Table \ref{tab:Jbar} and main text for more
    details). A typical value of the annihilation cross-section for
    thermally-produced DM is also plotted.}
  \label{fig:sfeldlimits}
\end{figure}

In the annihilation of dark matter particles to charged final states, internal
bremsstrahlung processes can contribute significantly to the high-energy end of the
\gr spectrum~\citep{2005PhRvL..95x1301B,Bringmann:2007nk}. Adding this
effect to the continuous spectrum of secondary $\gr$s from pion decay,
the total spectrum is given by
\begin{equation}
\frac{dN_{\gamma}}{dE_{\gamma}} = \frac{dN_{\gamma}^{\textrm{sec}}}{dE_{\gamma}} + \frac{dN_{\gamma}^{\rm IB}}{dE_{\gamma}} \;.
\end{equation}
The magnitude of this effect depends on the intrinsic properties of the dark
matter particle. \cite{Bringmann:2007nk} provide an approximation that
is valid for wino-like neutralinos \citep{2000NuPhB.570..455M}. The
annihilation spectrum for a 1 TeV wino is shown in
Figure~\ref{fig:spectra}. This parametrization is used in the calculation of the $95\%$~C.L. upper limit
on the velocity-weighted annihilation cross-section as a function of the
DM particle mass, presented in Figures ~\ref{fig:exclusionsAll}
and \ref{fig:sfeldlimits}. The internal bremsstrahlung affects the exclusion limits mostly in the low mass
DM particle regime, where its contribution to the total number of \grs in the H.E.S.S. acceptance is largest.

\subsection{\textit{Leptophilic} models}

Recent measurements of cosmic electron and positron spectra by
PAMELA~\citep{Adriani:2008zr}, ATIC~\citep{Chang:2008zzr}, H.E.S.S.~\citep{Aharonian:2009ah}
and \emph{Fermi}-LAT~\citep{Ackermann:2010ij} have been explained in terms of DM
annihilation primarily into leptonic final states (to avoid an
over-production of anti-protons), hereafter referred to as
\textit{leptophilic} models. \cite{2009PhRvL.103c1103B} show that the
\emph{Fermi}-LAT electron spectrum and the PAMELA excess in positron data can
be well explained by annihilation purely into $\mu^+ \mu^-$ pairs. In
this scenario, \grs are expected from final
state radiation (FSR) of the $\mu^+ \mu^-$ pair. While this final
state is rarely found in supersymmetric
models~\citep{1996PhR...267..195J}, some particle physics models
predict the annihilation to occur predominantly to lepton final
states~\citep{ArkaniHamed:2008qn,Nomura:2008ru}. The subsequent muon
decay into positrons and electrons may lead to an additional \gr
emission component by Inverse Compton (IC) up-scattering of background
photons,  such as those of the cosmic microwave background (CMB). If
the electron/positron energy loss time scale
is much shorter than the spatial diffusion time scale, the IC
contribution to the \gr flux may be
significant. In galaxy clusters, the energy loss term is dominated by
the IC component~\citep{2006AA...455...21C}. The total \gr spectrum
is then given by
\begin{equation}
\frac{dN_{\gamma}}{dE_{\gamma}} = \frac{dN_{\gamma}^{\rm FSR}}{dE_{\gamma}} + \frac{dN_{\gamma}^{\rm IC}}{dE_{\gamma}} \;.
\end{equation}
After extracting the FSR parametrization  from
\cite{2009PhRvD..79h3539B}, the IC component of the annihilation
spectrum was calculated following the method described in
\cite{2009JCAP...07..020P}. The total annihilation spectrum for a 1
TeV  dark matter particle annihilating to $\mu^+ \mu^-$ pairs is shown
in Figure~\ref{fig:spectra}. The energy E$^{\rm IC}_{\gamma}$ of the
IC emission peak is driven by electrons/positrons
of energy E$_{\rm e} \sim  m_{\rm DM}/2$ up-scattering target photons in a  radiation field of
average energy $\epsilon=$ 2.73 K and is given by $E^{\rm IC}_{\gamma} \approx
\epsilon(E_{\rm e}/m_{\rm e})^2$
\citep{2010hea..book.....L}. Consequently, the enhancement of the \gr
flux in the H.E.S.S. energy range is found to lower
the exclusion limits only for very high DM masses, $m_{\rm DM}
> 10$ TeV. The limits are enhanced by a factor of $\sim$10. The \emph{Fermi}-LAT exclusion limit for Fornax
is added (gray dashed-line), extending up to 10~TeV~\citep{2010JCAP...05..025A}. Due to the IC component, below a few tens of TeV the \emph{Fermi}-LAT results provide stronger limits than the H.E.S.S. results. However, since for DM particle masses above 10~TeV the IC emission peak falls out of the \emph{Fermi}-LAT energy acceptance, the IC spectra becomes harder in the same energy range. The \emph{Fermi}-LAT limits for DM particle masses above 10 TeV would tend to raise with a stronger slope than the slope in between 1 and 10 TeV. Thus H.E.S.S. limits would well-complement the \emph{Fermi}-LAT constraints in the DM mass range higher than 10~TeV.

\grs from IC emission are also expected in the case of DM particles annihilating purely into $b\overline{b}$. In the H.E.S.S. energy range for high DM masses ($\gtrsim$ 10 TeV) annihilating in the $b\overline{b}$ channel, the expected number of \grs including IC emission is lower than in the $\mu^+ \mu^-$ channel \citep[see, for instance,][]{Cirelli:2010xx}. This qualitative estimate in the \emph{Fermi}-LAT energy range (80 MeV - 300 GeV) shows that the number of expected \grs including IC emission for DM particle masses between 1 and 10 TeV is lower in the $b\overline{b}$ than in the $\mu^+ \mu^-$ channel by at least a factor of 2. Since the $\left\langle \sigma v \right\rangle$ exclusion limits are roughly scaled by the number of expected \grs, a qualitative estimate of the \emph{Fermi}-LAT limits including the IC component in the $b\overline{b}$ channel should not be better than their limits in the $\mu^+ \mu^-$ channel.

\subsection{\textit{Sommerfeld} enhancement}

The self-annihilation cross-section of dark matter particles can be
enhanced with respect to its value  $\sigmav _{0}$ during thermal
freeze-out by the \textit{Sommerfeld effect}~\citep[see e.g.][]{2004PhRvL..92c1303H,Profumo:2005xd}.
This is a velocity-dependent quantum mechanical effect:
If the relative velocity of two annihilating
particles is sufficiently low, the effective annihilation
cross-section can be boosted by multiple exchange of the force carrier bosons.
This can be parametrized by a boost factor $S$, as defined by:
\begin{equation}
\sigmav _{\rm eff} = S \times  \sigmav _{0} \, .
\label{eq:sfeld}
\end{equation}
\cite{PhysRevD.79.083523} consider the case of a
Sommerfeld boost due to the weak force which can
arise if the dark matter particle is a wino-like
neutralino. As a result of the masses and
couplings of the weak gauge bosons, the boost is strongest for a DM
particle mass of about 4.5 TeV, with resonance-like features appearing
for higher masses. This effect was proposed to account for the
PAMELA/ATIC data excess, where a boost of 10$^4$ or more is required
for neutralinos with masses of 1--10 TeV~\citep{Cirelli:2008pk}. It
was shown that the boost would be maximal in the dwarf galaxies and in
their substructures \citep{2009MNRAS.399.2033P}, due to the low DM
particle velocity dispersion in these objects.

In the Fornax galaxy cluster, the velocity dispersion and hence the mean
relative velocity of ``test masses'' such as stars, globular clusters or
galaxies is of the order of a few 100 km
s$^{-1}$~\citep{2010AA...513A..52S}, hence $\beta = \frac{<v_{\rm rel}>}{c} \approx
10^{-3}$. Assuming that the same velocity distribution holds true for
DM particles, limits on $\sigmav _{\rm eff}/S$ were derived which are
shown in Figure \ref{fig:sfeldlimits} for a signal integration radius
of $1.0\degr$ and the RB02 NFW profile. Although the DM velocity
dispersion is about one order of magnitude higher than in
dwarf galaxies, a boost of $\sim$10$^3$ is obtained for DM
particle masses around 4.5 TeV. The resonance-like feature is clearly
visible for masses above 4.5 TeV. Outside the
resonances, the limits on $\sigmav _{\rm eff}/S$ are tightened by more
than one order of magnitude for dark matter particles heavier than
about 3 TeV.

\subsection{Enhancement from dark matter substructures}

\begin{figure}[hb]
  \centering
  \includegraphics[width=\textwidth]{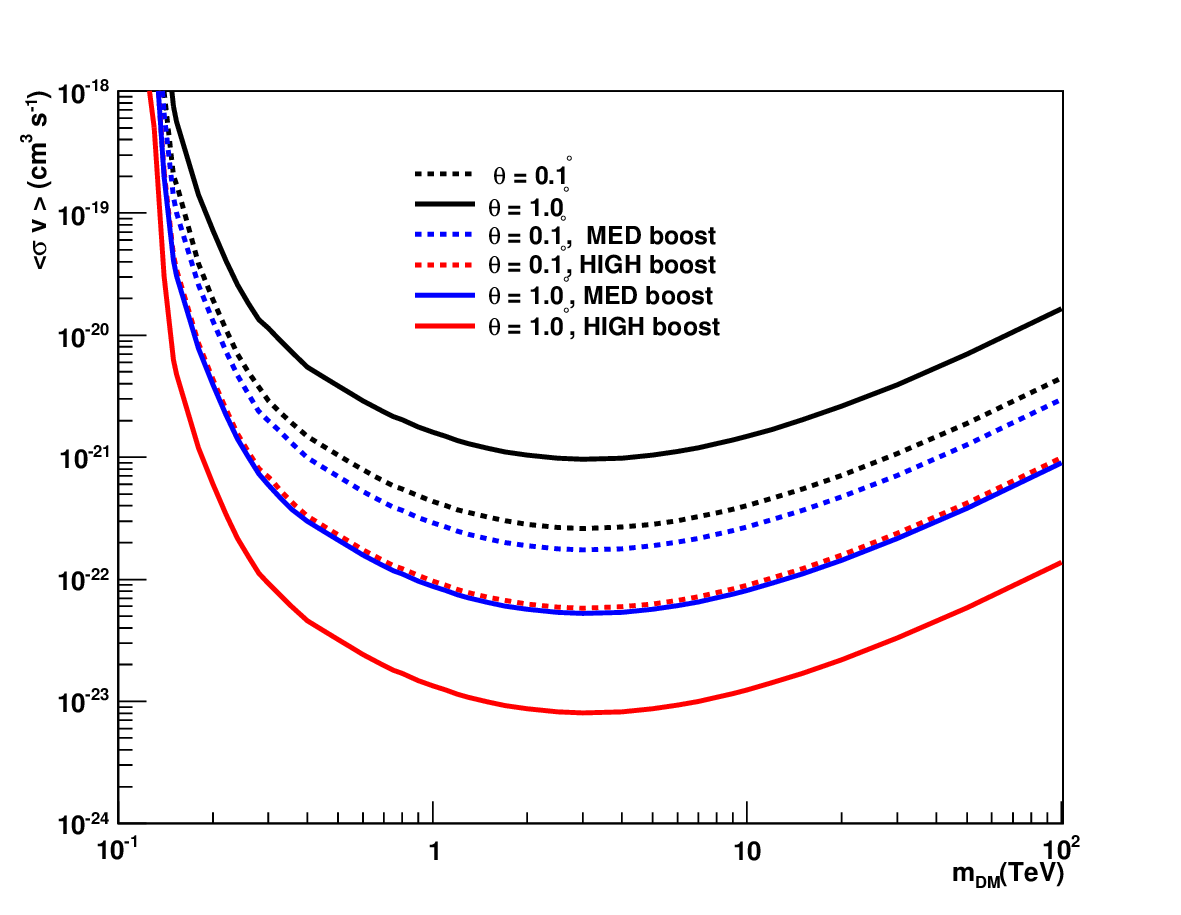}
  \caption{The effect of DM halo substructures:
    Upper limit at $95\%$ C.L. on \sigmav as function of the DM
    particle mass annihilating purely into $b\overline{b}$ pairs. The limits  are given for $\theta_{\rm max} =
    0.1\degr$  (dashed lines) and     $\theta_{\rm max} = 1.0\degr$
    (solid lines). The DM halo model RB02 is used (see Table \ref{tab:Jbar} and main text for more
    details). In addition, the effect of halo substructures
    on the \sigmav limits is plotted. The ``medium
    boost'' (MED) with $M_{\rm lim} = 5\,\times\,10^{-3}\;  \msun$  (blue lines)
    and the ``high boost'' (HIGH) with $M_{\rm lim} = 10^{-6}\; \msun$ (red lines)
    are considered.}
  \label{fig:exclusions4}
\end{figure}

The effect of DM substructures inside the opening angle of 0.1$^{\circ}$ and
1.0$^{\circ}$ are presented in Figure \ref{fig:exclusions4}, using the
enhancement values calculated in Section~\ref{sec:subs}. The
enhancements to the $95\%$ C.L. upper limits on $\left\langle \sigma v
\right\rangle$ are estimated using the two limiting masses of
substructures $M_{\rm lim}$. In the TeV range, the upper limit on
$\left\langle \sigma v \right\rangle$ is at the
$10^{-23}$~cm$^{3}$s$^{-1}$ level. The joint enhancement due to the
Sommerfeld effect added to the IB and the substructures contribution
is plotted in Figure \ref{fig:sfeldlimits}. In the most
optimistic model, with the largest enhancement by substructures and
the Sommerfeld effect, the $95\%$ C.L. upper limit on
$\sigmav _{\rm eff}$ reaches $10^{-26}$~cm$^{3}$s$^{-1}$, thus probing natural values for
thermally-produced DM.

\section{Summary}

The Fornax galaxy cluster was observed with the \hess telescope array to search for VHE
$\gr$s from dark matter self-annihilation. No significant \gr signal was
found and upper limits on the \gr flux were derived for power-law and DM spectra, at the level of 10$^{-12}$ cm$^{-2}$s$^{-1}$ above 260\gev.

Assuming several different models of particle dark matter and using published
models of the dark matter density distribution in the halo, exclusion limits on the DM
 self-annihilation cross-section as a function of the DM
particle mass were derived. Particular consideration was given
to possible enhancements of the expected \gr flux which could be caused
by DM halo substructure or the Sommerfeld effect. For a DM mass of 1 TeV,
the exclusion limits reach values of $ \sigmav \approx
10^{-22} - 10^{-23}$~cm$^{3}$s$^{-1}$, depending on DM
model and halo properties, without the substructures contribution, and $ \sigmav \approx
10^{-23} - 10^{-24}$~cm$^{3}$s$^{-1}$ when considering the substructures signal enhancement. At $M_{\rm DM} \approx 4.5$ TeV, a possible Sommerfeld resonance could lower the upper limit to $10^{-26}$~cm$^{3}$s$^{-1}$.

Compared to observations of dwarf spheroidal galaxies~\citep[see for instance][]{ScuCar} or globular clusters~\citep{Abramowski:2011hh}, these limits reach roughly the same order of magnitude. The choice of different tracers to derive the DM halo profile in Fornax galaxy cluster allows to well constraint the uncertainty in the expected signal. The  poorly constrained, but plausibly stronger subhalo enhancement in the Fornax cluster induces a uncertainty in the expected signal of about two orders of magnitude.

With an optimistic joint \gr signal enhancement by halo substructures and the Sommerfeld effect, the limits on \sigmav reach the values predicted for thermal relic dark matter. Additionally, they extend the exclusions calculated from \emph{Fermi}-LAT observations of galaxy clusters to higher DM particle masses.

\acknowledgments
The support of the Namibian authorities and of the University of
Namibia in facilitating the construction and operation of H.E.S.S.
is gratefully acknowledged, as is the support by the German
Ministry for Education and Research (BMBF), the Max Planck
Society, the French Ministry for Research, the CNRS-IN2P3 and the
Astroparticle Interdisciplinary Programme of the CNRS, the U.K.
Particle Physics and Astronomy Research Council (PPARC), the IPNP
of the Charles University, the South African Department of Science
and Technology and National Research Foundation, and by the
University of Namibia. We appreciate the excellent work of the
technical support staff in Berlin, Durham, Hamburg, Heidelberg,
Palaiseau, Paris, Saclay, and in Namibia in the construction and
operation of the equipment.

\clearpage

\bibliography{Fornax}



\end{document}